# A pharmacokinetic – viral kinetic model describes the effect of alisporivir monotherapy or in combination with peg-IFN on hepatitis C virologic response

Thi Huyen Tram Nguyen[1,2], France Mentré[1,2], Micha Levi[3], Jing Yu[4], Jérémie Guedj[1,2]

[1] IAME, UMR 1137, INSERM, F-75018 Paris, France

[2] IAME, UMR 1137, Univ Paris Diderot, Sorbonne Paris Cité, F-75018 Paris, France

[3] Novartis Pharmaceutical Corp., East Hanover NJ 07936, USA

[4] Novartis Institutes for BioMedical Research, Inc, Cambridge MA 02139, USA

**Abstract**

Alisporivir is a cyclophilin inhibitor with demonstrated in vitro and in vivo activity against hepatitis C virus (HCV). We estimated antiviral effectiveness of alisporivir alone or in combination with pegylated-Inteferon (peg-IFN) in 88 patients infected with different HCV genotypes treated for four weeks. The pharmacokinetics of both drugs were modeled and used as driving functions for the viral kinetic model. Genotype was found to significantly affect pegylated-Inteferon effectiveness ($\varepsilon$= 86.3% and 99.1% in genotype-1/4 and genotype-2/3, respectively, $p<10^{-7}$) and infected cells loss rate ($\delta$= 0.22 vs 0.39 day$^{-1}$ in genotype-1/4 and genotype-2/3, respectively, $p<10^{-6}$). Alisporivir effectiveness was not significantly different across genotype and was high for doses ≥600 mg QD. We simulated virologic responses with other alisporivir dosing regimens in HCV genotype-2/3 patients using the model. Our predictions consistently matched the observed responses, demonstrating that this model could be a useful tool for anticipating virologic response and optimize alisporivir-based therapies.

**Introduction**

Chronic infection with hepatitis C virus (HCV) affects about 170 million people worldwide and is responsible for the majority of morbidity and mortality from liver diseases.[1] The goal of treatment is to achieve a sustained virologic response (SVR), marker of viral eradication, assessed by the absence of detectable HCV RNA six months after treatment cessation. Over the last few years, a rapid development of direct acting antivirals (DAAs) has opened a new era for HCV therapy. The advent of the new therapies has dramatically increased SVR rate and decreased treatment duration. In recent phase 3 clinical trials including a protease or a polymerase inhibitor, SVR was higher than 90% following a 12 week-treatment. This holds the expectation that several one-in-a-pill (i.e., interferon-free) treatments with universal cure might be achieved.[2,3] However several challenges are still ahead. First DAAs have been optimized to treat HCV genotype-1 and have lower effectiveness against genotypes2/3, which account for 19 to 40% of HCV-infected patients. Second, high mutation rate of HCV of several DAA classes (i.e. protease and non-nucleoside polymerase inhibitors) requires therapies with a high barrier to resistance, which can be achieved by combining several antiviral agents.[2,4] However therapy combination can be a particularly difficult endeavor and requires anticipating possible pharmacokinetic-pharmacodynamic (PK-PD) interaction between different agents.

Unlike DAAs that specifically block a viral protein, host targeting antivirals (HTA) block host proteins necessary for viral replication.[2] Alisporivir, a second-generation cyclophilin inhibitor devoid of immunosuppressive activity, is the most advanced HTA.[2,5] Alisporivir acts primarily by blocking in a dose-dependent manner the interaction between cyclophilin A and the HCV NS5A protein.[5] It has shown potent anti-HCV activity across various HCV genotypes in several *in vitro* and *in vivo* studies.[5–7] Alisporivir was also found to have a high genetic barrier to resistance.[8] In a phase 2a study (DEB-025-HCV-203, Euract number: 2006-002695-17), the virologic response to alisporivir monotherapy or in combination with pegylated interferon (peg-IFN) was studied in 90 patients infected with HCV genotype 1, 2, 3 or 4.[9] Alisporivir (200, 600, and 1000 mg given BID for one week then QD till the end of treatment) showed a dose-dependent antiviral effect in combination with peg-IFN, with mean viral decay of more than 4 and 5 $\log_{10}$IU/mL at week 4 in HCV genotype1 and genotype-2/3, respectively.[9] However, multiple factors involved (genotypes, monotherapy vs combination, alisporivir doses) made it difficult to anticipate the effect of different alisporivir dosing regimen on longer time period.

Mathematical models for HCV kinetics have provided important insights into HCV life cycle as well as the effectiveness and the mechanisms of action of different anti-HCV agents.[10–14] The understanding of virologic response can be further improved by incorporating PK information.[11,15] However this approach became more complicated in drug combination as both PK and PD interaction can occur. Here we incorporated the PK of alisporivir and peg-IFN into a viral kinetic (VK) model to estimate the antiviral effectiveness of both agents and to tease out the effect of genotypes and alisporivir doses on the virologic response. Lastly we evaluated the robustness of our model by comparing the predictions of the model and virologic responses observed in a phase 2b study (VITAL-1, clinicaltrials.gov ID: NCT01215643), where alisporivir was given at different dosing regimens with response-adapted treatments.

**Results**

**Data**

The study population comprised 90 HCV patients, 57.8% were men and 66.7% were infected with HCV genotype-1. Baseline and patients characteristics are presented in Table S1.[9]

This was a randomized, double-blind, placebo-controlled phase 2 study in treatment naïve HCV patients with compensated liver function. Patients were randomized into 5 treatment arms receiving peg-IFNα2a (180 μg/week) combined with placebo, 200, 600 or 1000 mg alisporivir or 1000 mg alisporivir monotherapy (Arm PEG, A, B, C, D, respectively). Alisporivir was administered as twice daily (BID) for one week, followed by once daily (QD) for three weeks. No lack of compliance was reported. Alisporivir and peg-IFN plasma trough concentrations were determined on days 1, 8, 15, 22, 29, and 50 (21 days after the last dose). In addition, a 12-hour and 24-hour PK of alisporivir and peg-IFN was established on days 1 and 29 in a subset of patients. Plasma concentrations of alisporivir and peg-IFN were determined using validated LC/MS/MS and ELISA assays with the limit of quantification (LOQ) of 2.5 ng/mL and 0.17 ng/mL, respectively.[16] HCV RNA levels (IU/mL) were determined by the Roche COBAS TaqMan, Roche laboratories, Basel, Switzerland with LOQ of 15 or 45 IU/mL and detection limits of 10 or 15 IU/mL.

The PK model of alisporivir was developed using 858 observations from 50 patients with rich design and 104 observations from 21 patients with sparse sampling. Seven post-dose concentrations recorded as trough concentrations were excluded from the analysis.

The PK model of peg-IFN was developed using 527 observations from 48 patients with rich design and 100 observations from 21 patients with sparse design.

The VK model was developed using on- and post-treatment data (521 viral loads from 88 patients). Two patients having no PK record for both alisporivir and peg-IFN were excluded. The percentages of viral load below quantification limits were 27.6%.

**Alisporivir PK model**

Alisporivir PK on day 29 was described well using a two-compartment model with delayed first-order absorption and Michaelis-Menten elimination (Table S2). However, this model over predicted alisporivir PK profile on day 1 (Figure S1). This discrepancy between day 1 and day 29 PK was described using a time-dependent bioavailability. The structural model is described as below:

$$F(t) = F_0 + (F_{inf} - F_0)(1 - \exp(-k_F t))$$

$$\frac{dA}{dt} = \begin{cases} 0 & \text{if } t \leq T_{lag} \\ -k_a A & \text{if } t > T_{lag} \end{cases}$$

$$\frac{dQ_1}{dt} = -\frac{dA}{dt} - k_{12}Q_1 + k_{21}Q_2 - \frac{V_m Q_1}{K_m V + Q_1}$$

$$\frac{dQ_2}{dt} = k_{12}Q_1 - k_{21}Q_2$$

where F(t) is the bioavailability at time t. A, $Q_1$, $Q_2$ are the drug quantity at the absorption site, in central and peripheral compartment, respectively. As the bioavailability could not be identified using only oral data, we fixed $F_0$ at the value found in monkeys (0.27). The relative bioavailability reached its maximal value in approximately 4 days ($k_f$). The maximal relative bioavailability ($F_{inf}$) was approximately 3-fold higher as compared to the base line value ($F_0$). We also tested time-decreasing $V_m$ or time-increasing $K_m$ models but their Bayesian Information Criterion (BIC), a criterion for model selection based on the log-likelihood penalized by number of parameters and observations (the lower the better), were much higher, compared to time-dependent bioavailability (Table S3).

However, alisporivir exposure following 1000 mg QD was higher at day 29 in the combination therapy as compared to monotherapy. In order to describe this higher exposure we tested a possible peg-IFN effect on alisporivir PK (peg-IFN effect on $V_m$, $K_m$ and $F_{inf}$). A statistically significant effect of peg-IFN on reducing alisporivir elimination at the dose of 1000 mg QD was described by decreasing both $V_m$ (55 vs 115 mg.h$^{-1}$, p<0.003) and $K_m$ (0.64 vs 2.44 mg.L$^{-1}$, p<0.00023). Peg-IFN effect on the $K_m$ and $V_m$ led to 1.2-time increase in the median value (or 1.5-time increase in the mean value) of AUC on day 29. The parameter estimates for alisporivir PK were shown in Table 1.

**Peg-IFN PK model**

The PK of peg-IFN was described by one-compartment model with linear absorption and elimination. alisporivir had no significant effect on peg-IFN PK (Table 1).

**PK-VK model**

The shrinkages of PK individual parameters (computed from ratio of variances) were higher than 50% for some PK parameters of alisporivir such as $k_a$ (58%), $k_{12}$ (77%), $V_m$ (58%), $K_m$ (51%) and $F_{inf}$ (60%) as there were 21 out of 71 patients having only weekly assessed trough concentrations. However, using individual PK profiles predicted from the individual parameters, the PK-VK model fitted to the viral load data very well and was able to describe the fluctuation in virologic response due to the variation of concentrations over the dosing period (Figure 3).

HCV genotypes were found to significantly affect both the effectiveness of peg-IFN and the loss rate of infected cells but not the effectiveness of alisporivir (Table 1). The average effectiveness of peg-IFN in the seven days following the last dose was higher in genotype-2/3 than in genotype-1/4 patients with median values of 99.1% and 86.3%, respectively ($p<10^{-8}$). Likewise the loss rate of infected cells, $\delta$, was higher in HCV genotype-2/3 than in genotype-1/4 patients (0.39 day$^{-1}$ vs 0.22 day$^{-1}$, respectively, $p<10^{-6}$). Alisporivir effectiveness reached the highest values at the end of BID dosing during week 1 with a median value of 71.3%, 95.2%, 97.6% and 95.7% in arms A, B, C and D (alisporivir 200 mg, 600 mg, 1000 mg with peg-IFN and alisporivir 1000 mg monotherapy), respectively and then dropped when switching to QD dosing with a median effectiveness at the last day of treatment of 50.5%, 89.4%, 97.1% and 93.6% in arm A, B and D, respectively. Higher alisporivir effectiveness in combination than in monotherapy (median $\varepsilon$=97.6% vs 95.7% at day 8 and 97.1% vs 93.6% at day 29, respectively) was explained by higher exposure of alisporivir due to PK interaction between the two drugs. Given in combination, alisporivir and peg-IFN acted additively and yielded to high effectiveness, with mean effectiveness at day 29 in HCV genotype-2/3 patients equal to 98.5%, 99.3% and 99.5% in arm A, B and C, respectively as compared to 92.0%, 96.1% and 99.3% in genotype-1/4 patients in arm A, B and C, respectively. The VK in different treatment arms and in two genotype populations was illustrated in Figure 4.

**External validation**

We used the PK-VK model to simulate virologic responses and compared them to the results from the VITAL-1 study (see Methods). The observed responses were included within the 95% prediction interval in the large majority of cases evaluated (Table 2). Likewise a good match was also obtained for later data points (after week 6) where treatment regimen depended on the initial virologic response.

**Discussion**

Models of HCV kinetics have become valuable tools to estimate antiviral effectiveness and optimize therapy.[10,17–19] Initially developed for peg-IFN/ribavirin they have been recently applied to various classes of DAA agents.[10,12] However most of the results published considered constant treatment effectiveness over time and/or empirical function to describe varying-effectiveness models.[11,12,15,20] In fact there have been only a few attempts to include PK into VK models, even less in drug combinations.[21] This, indeed, considerably complicates the analysis as interactions are susceptible to occur both at the PK and PD levels. Here we characterized the VK obtained in a phase 2a study using both alisporivir and peg-IFN PK information.

Alisporivir PK model was developed via several steps. First a structural model using day 29 data was developed. A two-compartment model with Michaelis-Menten elimination was the best to describe these data, consistent with the nonlinearity in alisporivir PK reported in previous studies.[22,16] This model (based on day 29 data) systematically over-estimated the concentrations of alisporivir after the first dose on day 1. This overestimation could be due to a time-varying PK or to non-linearity in absorption and/or distribution and/or elimination that was not yet properly handled. To correctly identify the cause for this overestimation and to build a more mechanistic PK model for alisporivir, further data such as IV data and more detailed PK profiles during treatment would be required. We modeled the discrepancy of PK between day 1 and 29, using an empirical model allowing PK parameters ($V_m$, $K_m$ and bioavailability) to change with time. The model assuming time-dependent bioavailability was the best statistical model to describe the full PK profile. Here we chose to model the change in exposure, which may be the consequences of changes in bioavailability and/or distribution volume, using a time-varying bioavailability. The relevance of this model can be explained by the fact that as alisporivir concentration increases, the transporters involved in the absorption and tissue distribution of alisporivir (P-gp, MRP2, OATP1B1, OATP1B3, NTCP, BSEP) could be inhibited.[7] This would increase alisporivir plasma concentration (by increasing bioavailability and reducing the uptake to the liver and other distribution sites) and hence, increase bioavailability.

The effect of peg-IFN on alisporivir PK was investigated by adding peg-IFN as a covariate on alisporivir PK parameters. Peg-IFN decreases both $V_m$ and $K_m$. Indeed, IFN was shown to affect the activity and the expression of the cytochrome P450 system[23–27] and of several drug transporters such as P-gp, MRP2, MRP3, BCRP, BSEP in human hepatocytes.[27–29] These proteins are all involved in the elimination of alisporivir, via metabolism (mainly by CYP3A4) or excretion via the bile as parent drug. Although the mechanism for the PK interaction between alisporivir and peg-IFN remains unclear, it could be in part explained by the inhibition effect of peg-IFN on the CYP450 system, possibly on CYP3A, and on drug transporters such as P-gp. This hypothesis is also supported by the case of telaprevir, another substrate of CYP3A and P-gp[30], where plasma exposure tends to increase in presence of peg-IFN.[31,32] However, this interaction was evaluated using only an alisporivir dose of 1000 mg and additional monotherapy dosing cohorts are needed to evaluate whether this effect is universal or specific to this dosing regimen. Also, in this analysis, we did not account for the fluctuation in peg-IFN exposure that may influence the drug-drug interaction.

Peg-IFN PK is well described with a one-compartment model. The parameters obtained were close to what had been reported during peg-IFN/ribavirin therapy.[33–35]

We used estimated individual parameters to predict individual PK profiles of both drugs and these predictions were incorporated into a VK model to fit on- and post-treatment viral load data. A more rigorous approach could be to simultaneously fit PK and VK. However the sequential approach has been shown to lead to limited bias while being much less time-consuming than simultaneous approach.[36]

To examine if the model can explain well the VK in each treatment arm, we used plots of prediction discrepancies versus time. Prediction discrepancies are simulation-based evaluation metrics, similar to visual predictive check (VPC) but unlike VPC, the heterogeneity in covariate values are handled in their computation [37–39]. Therefore, we can use one prediction discrepancies plot for all data from all genotypes instead of splitting data into genotype-1/4 and 2/3 groups (with 12 and 6 patients, respectively) to make two VPC plots for each treatment arm. Assuming a Bliss independence model for the combined effect, we obtained satisfactory prediction discrepancies plots for all treatment arms (Figure 5), which indicated that this model without PD interaction was sufficient to characterize the combined effect. Although a slightly synergistic effect was observed in vitro,[40] the data of this study did not allow ruling out any significant synergy.

With this PK-VK model, the improved virologic response observed in HCV genotype-2/3 patients receiving dual-therapies was attributed to a larger loss rate of infected cells ($\delta = 0.39$ day$^{-1}$ vs 0.22 day$^{-1}$ in genotype-2/3 and genotype-1/4, respectively, $p=10^{-6}$) and a lower IC50$_{peg-IFN}$ than in genotype-1/4 patients (0.146 µg.L$^{-1}$ vs 2.26 µg.L$^{-1}$, corresponding to 99.1% vs 86.3% for $\varepsilon_{peg-IFN}$, $p=10^{-8}$). These results are consistent with previous studies comparing VK parameters between HCV genotype-2/3 and genotype-1/4 patients with peg-IFN/ribavirin.[41,42] Our analysis revealed that the loss rate of infected cells in the alisporivir monotherapy arm was also larger in genotype-2/3 patients ($\delta=0.41$ in

genotype-2/3 vs 0.23 day$^{-1}$ in genotype-1/4 patients, respectively, $p<10^{-11}$), supporting that larger loss rate of infected cells was probably inherent to HCV genotypes and independent from the drug. In contrast, we could not show any significant effect of HCV genotypes on alisporivir effectiveness; however this effect was associated with a large confidence interval. Further studies with an appropriate design allowing better estimation of antiviral effectiveness, i.e., with frequent sampling in the first 3 days of treatment,[43] are needed to correctly evaluate this effect. Altogether our analysis suggested that the larger response observed in HCV genotype-2/3 receiving alisporivir monotherapy was due to an intrinsically larger loss rate of infected cells and not to a differential effect of alisporivir in blocking virus production.

Next, in order to validate our model on an independent dataset, we used our model to predict the early virologic response (until week 12) with the design of the VITAL-1 study, where alisporivir was given in combination with ribavirin or peg-IFN for 24 weeks. To do this, two main simplifications were made. First, the effect of ribavirin was neglected, consistent with the hypothesis that ribavirin primarily acts against drug-resistant virus and has no or low effect on drug-sensitive virus.[44,45] Second, our model was built on short term data until day 29 where no resistance-related viral breakthrough was observed. Therefore we excluded from our analysis patients treated with alisporivir monotherapy without ribavirin where significant prevalence of non-PK related suboptimal response (18 among 83 patients, corresponding to 21.5%)[22] was reported which requires to incorporate other mechanisms such as resistance.

Therefore three dosing regimens were considered, two with alisporivir/ribavirin (600 or 800 mg QD) and one with alisporivir/peg-IFN (600 mg QD). The prediction was complicated by the response-adapted treatment strategy. Interestingly, our predicted proportions of patients with undetectable virus matched very well the observed results in all groups, including the IFN-free regimen. A small discrepancy between predictions and observations was noticed at week 6, where we slightly overestimated the proportion of patients with undetectable HCV RNA in the group receiving alisporivir 800 mg without IFN (57.4% vs 45.0%). Nevertheless, the observed response was probably artefactually low, as hinted by the fact that it was 5 points less than that observed with alisporivir 600 mg without IFN. Our model also predicted slightly lower responses at week 12 for these two IFN-free arms (~93% vs 98% observed in VITAL-1) and this under-prediction could be explained by the fact that ribavirin could help to improve the second phase decline in patients whose responses to peg-IFN are low while in the simulation, we ignored this effect of ribavirin on add-on peg-IFN treatment.

Given the robustness of our early predictions, we extrapolated the proportion of patients considered as cured after 24 weeks of treatment, using a theoretical cure boundary. This cure boundary is a theoretical threshold for viral eradication, defined as less than one viral particle or infected cell remaining in 15L of extracellular fluid[46]. It was shown to provide good predictions for SVR in previous studies.[46,47] Here as well our predictions nicely matched reported results (Table 2). Our simulations also suggested that treatment duration could be optimized using early virologic response.

In conclusion, we introduced a PK-VK model for drug combination. This model could well describe VK observed with different dosing groups of alisporivir given alone or with peg-IFN. Moreover this model provided good predictions for both early and long-term response of another study. Overall our approach demonstrates the use of mathematical modeling to anticipate treatment outcome in the coming era of oral combinations therapy for chronic HCV infections.

**Methods**

**Alisporivir PK structural model**

We built the structural model using frequent sampling PK data at day 29 (steady-state) in patients receiving alisporivir/peg-IFN. For this, we tested one-, two-, three-compartment models, linear and Michaelis-Menten elimination, linear absorption with or without a latent time. We then included day 1 data and trough concentrations obtained in this subset of patients to characterize the change in PK over time. This was achieved by allowing the elimination parameters and the apparent bioavailability to change over time:

$$\theta(t) = \theta_0 - (\theta_{inf} - \theta_0)(\exp(-k_\theta t) - 1)$$

where $\theta(t)$ is value of the concerning parameter $\theta$ at time t, $\theta_0$ is the value at time 0, $\theta_{inf}$ is the maximum/minimum value achieved at infinity and $k_\theta$ is the changing rate.

**Peg-IFN PK structural model**

A one-compartment model was fitted to all the available data.[33–35]

**PK interaction**

Once the PK structural models of both alisporivir and peg-IFN were determined, the impact of alisporivir or peg-IFN on the other drug was evaluated by including alisporivir (yes/no), peg-IFN (yes/no) as a covariate.

**PK individual parameters**

After PK models for both drugs were obtained, individual parameters were retrieved as the Empirical Bayes Estimates (EBEs). The PK predictions obtained with these EBE and dosing information were used as driving functions in VK model.

**Viral kinetic modeling**

The following VK model was fitted to viral load data:

$$\frac{dI}{dt} = \beta VT - \delta I$$

$$\frac{dV}{dt} = (1-\varepsilon)pI - cV$$

where T, I and V represent target cells, infected cells and free virus, respectively. We assumed a constant liver volume (i.e., T+I) during the study period. Infected hepatocytes die at rate $\delta$ and release virions at a rate $p$ per cell. Virions infect target cells with infection rate $\beta$ and are cleared from serum with rate $c$. Anti-viral drugs are supposed to act mainly by blocking virion production, $p$, with effectiveness $\varepsilon$. Because c cannot be estimated when there is no viral load data during the first week of treatment, we fixed c = 6 day$^{-1}$ as done previously.[48]

Next the concentration effect relationship was modeled using an Emax model:

$$\varepsilon_{peg-IFN}(t) = \frac{C_{peg-IFN}(t)}{IC50_{peg-IFN} + C_{peg-IFN}(t)}$$

$$\varepsilon_{ALV}(t) = \frac{C_{ALV}(t)}{IC50_{ALV} + C_{ALV}(t)}$$

where IC50$_{ALV}$ and IC50$_{peg-IFN}$ are concentrations of alisporivir and peg-IFN, respectively, needed to achieve an effectiveness of 50%. C$_{ALV}$(t) and C$_{peg-IFN}$(t) are the drug concentrations predicted at time t.

Lastly the total effectiveness of alisporivir and peg-IFN was modeled using a Bliss independence equation to describe the combined effectiveness of two drugs having independent mechanisms of action[49]:

$$\varepsilon = 1 - (1-\varepsilon_{ALV})(1-\varepsilon_{peg-IFN})$$

The effect of HCV genotypes was studied by including the covariate genotype1/4 (yes/no) on all VK parameters and treatment effectiveness.

**Individual mean effectiveness**

VK individual parameters were retrieved as the Empirical Bayes Estimates (EBEs). Then, the individual average effectiveness over a dosing interval [t$_i$, t$_j$] is defined by:

$$\bar{\varepsilon}\Big|_{t_i}^{t_j} = \frac{\int_{t_i}^{t_j} \varepsilon(t)dt}{t_j - t_i}$$

**Statistical models**

Let f the function describing the structural model, the statistical model for the observation $O_{ij}$ (drug concentrations or $\log_{10}$ viral loads) of subjects i at time $t_{ij}$ is:

$$O_{ij} = f(\theta_i, t_{ij}) + e_{ij}$$

where $\theta_i$ is the vector of parameters of subject i and $e_{ij}$ is the residual error.

The individual parameters $\theta_i$ are supposed to follow log-normal distribution:

$$\theta_i = \mu \exp(\beta_{cov} \times COV_i) \exp(\eta_i)$$

or logit-normal distribution (for the bioavailability parameter):

$$\theta_i = \frac{\mu}{\mu + (1 - \mu) \exp(-\beta_{cov} \times COV_i) \exp(-\eta_i)}$$

where $\mu$ is the fixed effects, representing population values, $\beta_{cov}$ is the covariate effect, $COV_i$ is the binary covariate (treatment combination or genotype 1/4) and $\eta_i$ is the random effect, supposed to follow $N(0, \omega^2)$.

The residual error $e_{ij}$ were assumed to follow $N(0, \sigma^2_{ij})$, where $\sigma^2_{ij}$ can be additive ($\sigma_p = 0$), proportional ($\sigma_a = 0$) or combined model:

$$\sigma_{ij}^2 = \left(\sigma_a + \sigma_p f(\theta_i, t_{ij})\right)^2$$

We supposed a combined and an additive model for PK and VK error model, respectively.

**Parameter estimation**

Parameters were estimated using the SAEM algorithm implemented in MONOLIX 4.2.2 to handle data below LOQ. Models were compared using BIC values. Covariate was selected using stepwise method, based on p-value of likelihood ratio test. The likelihood was obtained using important sampling method. Only random effects correlations with a Pearson's coefficient higher than 0.7 were tested. Standard goodness-of-fit plots, visual predictive checks, prediction discrepancies[37–39] were used for model evaluation.

**External validation**

We evaluated the predictive ability of the PKVK model by comparing model predictions with the virologic responses obtained in patients genotype 2/3 in three alisporivir-containing arms of the VITAL-1 study.[5] All the three groups received a loading dose of 600 mg alisporivir BID for 1 week, followed by alisporivir 600 mg QD + Ribavirin (ribavirin) (arm VITAL-1b, N=84), alisporivir 800 mg QD + ribavirin (arm VITAL-1c, N=94) and alisporivir 600 mg QD + peg-IFN (arm VITAL-1d, N=39).

Patients with RVR, i.e., HCV RNA below LOQ (<25 IU/mL) at treatment week 4, continued the same treatment until week 24, while non-RVR patients switched to alisporivir 600 mg QD plus peg-IFN/ribavirin from week 6 to week 24. Of note, the VITAL-1 study also comprised a group treated with alisporivir monotherapy. But a large proportion of patients developed non-PK related suboptimal response which requires other mechanism (e.g. resistance) to be incorporated.[22] Thus this arm was excluded from our simulation.

We used the PK-VK model (including residual errors) to simulate the virologic responses in the three alisporivir-containing arms of VITAL-1, considering the adaptive treatment and the number of patients retained in per protocol analysis. We assumed 100% treatment adherence and considered that ribavirin could prevent the growth of resistant viruses but had no effect on drug sensitive viruses[44], i.e., we simulated alisporivir/ribavirin co-treatment using only the PKVK model of alisporivir. For each simulated patient, viral eradication (i.e., SVR) was considered attained if the model predicted less than 1 virion or infected cells in the whole extracellular fluid volume, i.e., $6.7 \times 10^{-5}$ copy(cell)/mL, during treatment period.[46]

**Study Highlights**

*What is the current knowledge on the topic?*

Alisporivir is a host-targeting agent against HCV currently in phase 3 of clinical development. In a phase 2 dose-ranging study, the combination of alisporivir/peg-IFN for 28 days led to a rapid virologic decline for various HCV genotypes.

*What question does this study address?*

We built a pharmacokinetic/viral kinetic model to estimate antiviral effectiveness of alisporivir and peg-IFN and predicted the virologic responses with other therapeutic strategies.

*What this study adds to our knowledge?*

This pharmacokinetic/viral kinetic model accounted for PK/PD interactions in drug combination in HCV. Alisporivir is highly effective at doses higher than 600 mg QD and acts additively with peg-IFN without evidence of synergism. Alisporivir effectiveness was not found to be significantly different across genotypes. Faster virologic response in genotype-2/3 patients was due to a more elevated loss rate of infected cells and higher sensitivity to peg-IFN.

*How this might change clinical pharmacology and therapeutics?*

This study showed that model from early viral kinetics could be useful to predict treatment outcome and anticipate future clinical studies.

**Conflict of Interest**

T.H.T.N. has a research grant from Novartis. J.Y. and M.L. are working for Novartis, in Advanced Quantitative Sciences department.

Author Contributions

T.H.T.N., J.G. and F.M. designed the simulation study. T.H.T.N., J.G. and F.M. participated in statistical analysis. T.H.T.N. performed the simulations. T.H.T.N., J.G. wrote the manuscript. All authors helped to draft the manuscript, read and approved the final manuscript.

**Acknowledgements**

The authors would like to thank Nikolai V. Naoumov for kindly providing the data. Comments and inputs given by Novartis HCV clinical team have been a great help in improving the manuscript. I would also like to thank Hervé Le Nagard and François Cohen for letting me with use the Fisher Machine for the simulation.

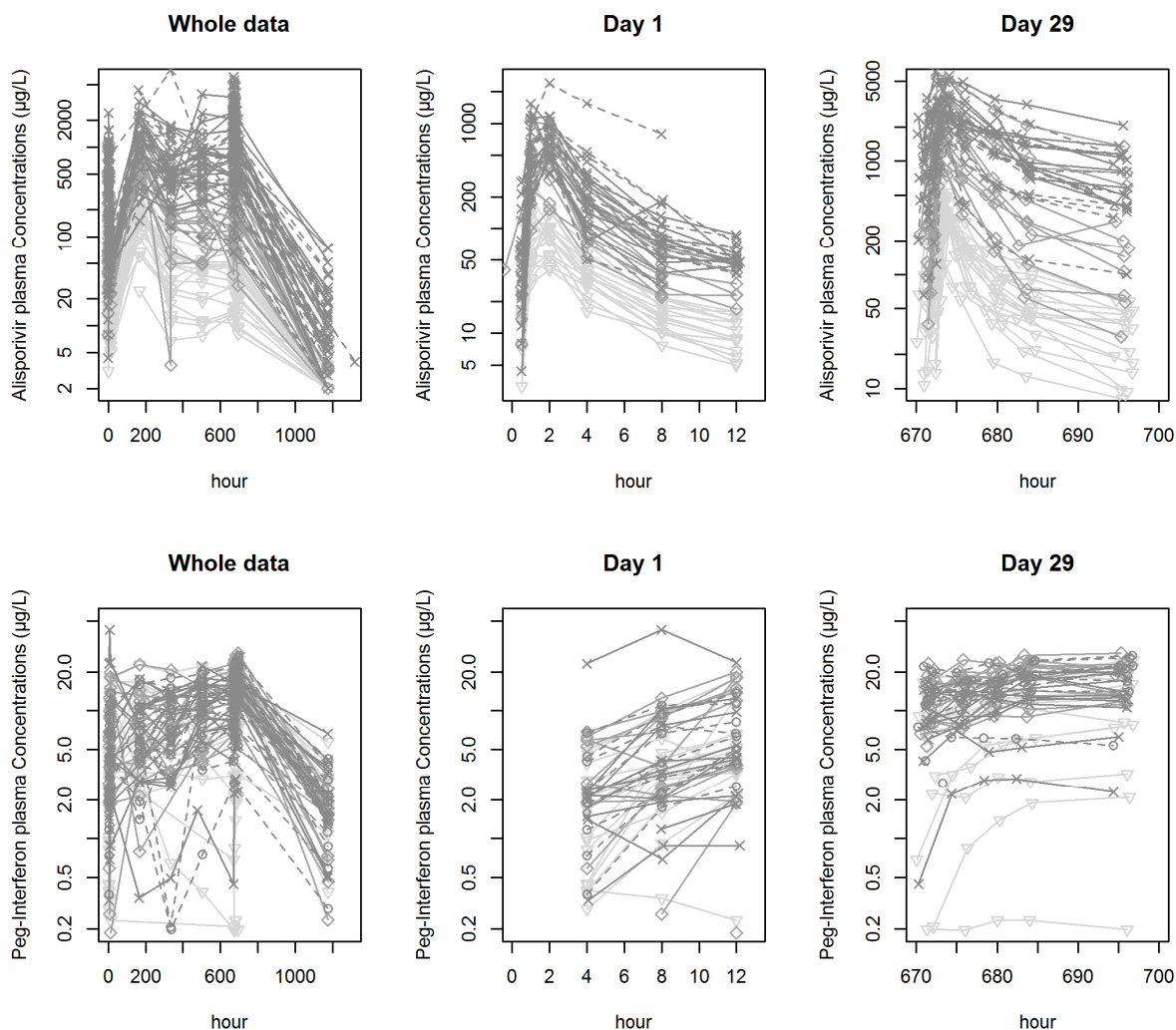

Figure 1. Spaghettiplots of drugs concentration over time.

Upper row: Spaghettiplots of alisporivir concentration over time in different treatment arms (Arm A - straight line and triangular, Arm B - straight line and diamond, Arm C - straight line and cross, Arm D - dashed line and cross) in the whole study period (left), on the first day (middle) and on the last day of treatment (right)

Last row: Spaghettiplots of peg-IFN concentration over time in different treatment arms (Arm A - straight line and triangular, Arm B - straight line and diamond, Arm C - straight line and cross, Arm PEG - dashed line and circle) in the whole study period (left), on the first day (middle) and on the last day of treatment (right)

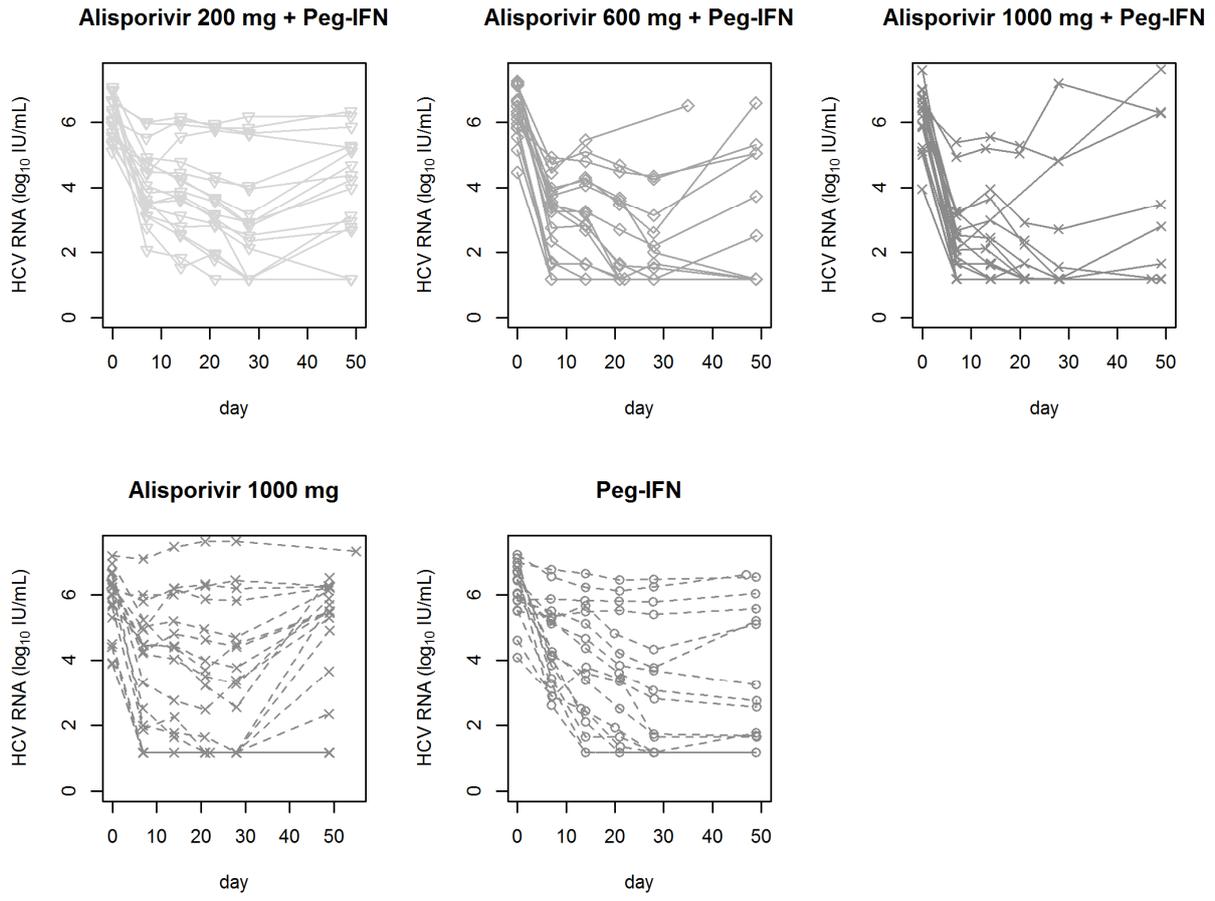

Figure 2. Spaghettiplots of viral load over time in different treatment arms (Arm A - straight line and triangular, Arm B - straight line and diamond, Arm C - straight line and cross, Arm D - dashed line and cross, Arm PEG - dashed line and circle) in the whole study period

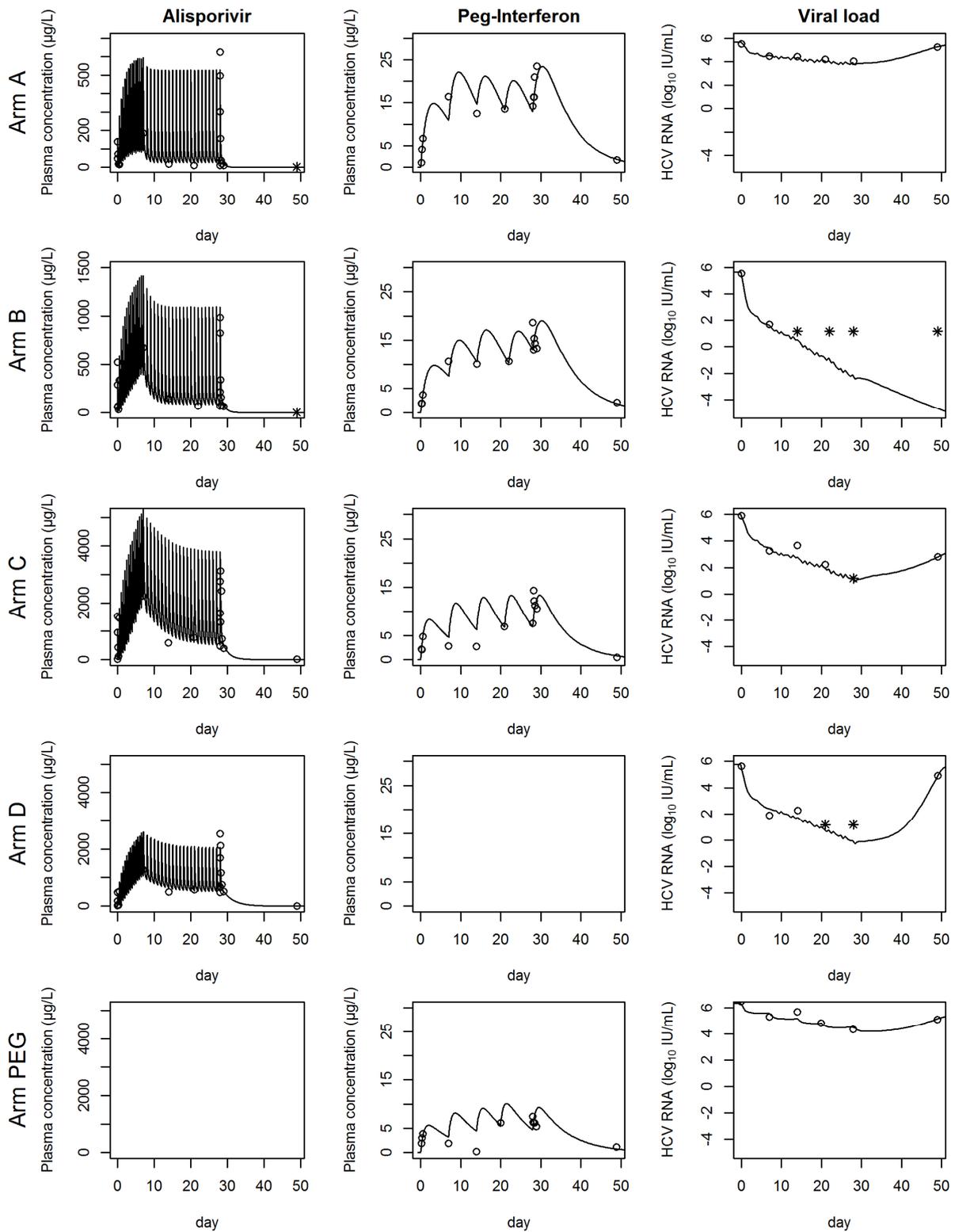

Figure 3. Some examples of individual fits for 5 treatment arms obtained with PK-VK model. Arm A (alisporivir 200 mg + peg-IFN 180μg): 1st row, Arm B (alisporivir 600 mg + peg-IFN 180μg): 2nd row, Arm C (alisporivir 1000 mg + peg-IFN 180μg): 3rd row, Arm D (alisporivir 1000 mg): 4th row, Arm Peg (peg-IFN 180 μg): last row. Fits for alisporivir PK, peg-IFN PK and viral load are presented on the 1st,

2nd and 3rd column, respectively. Observed data are presented as circle; data below the quantification limit are presented as star symbols.

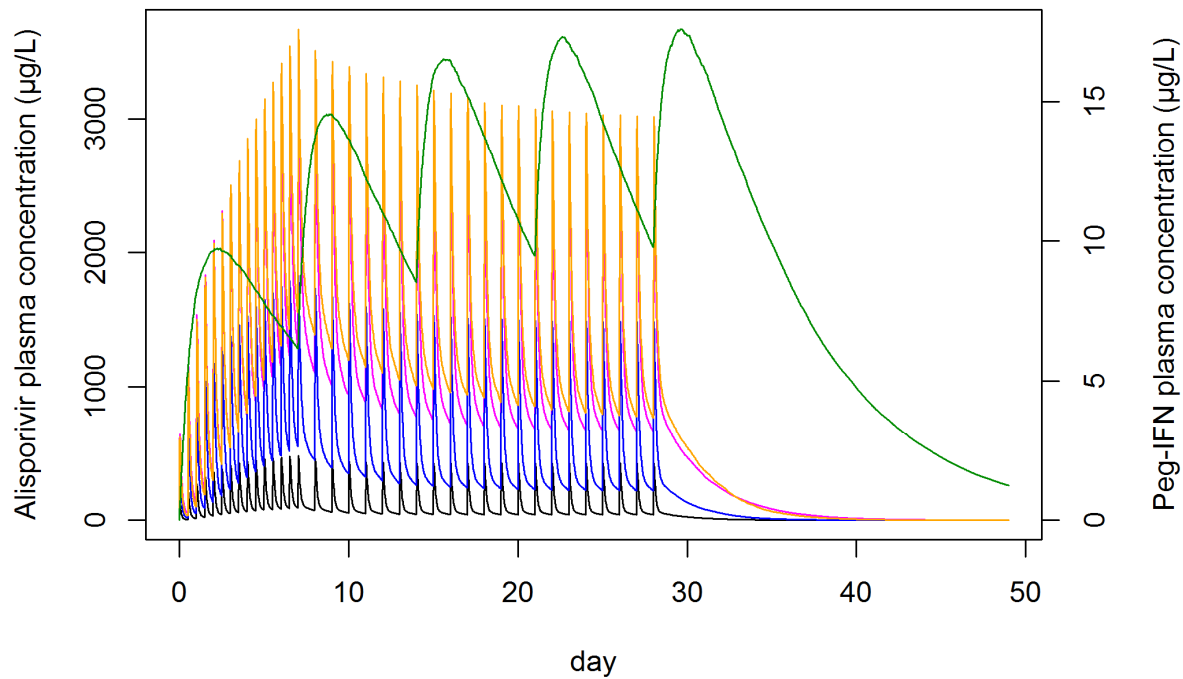

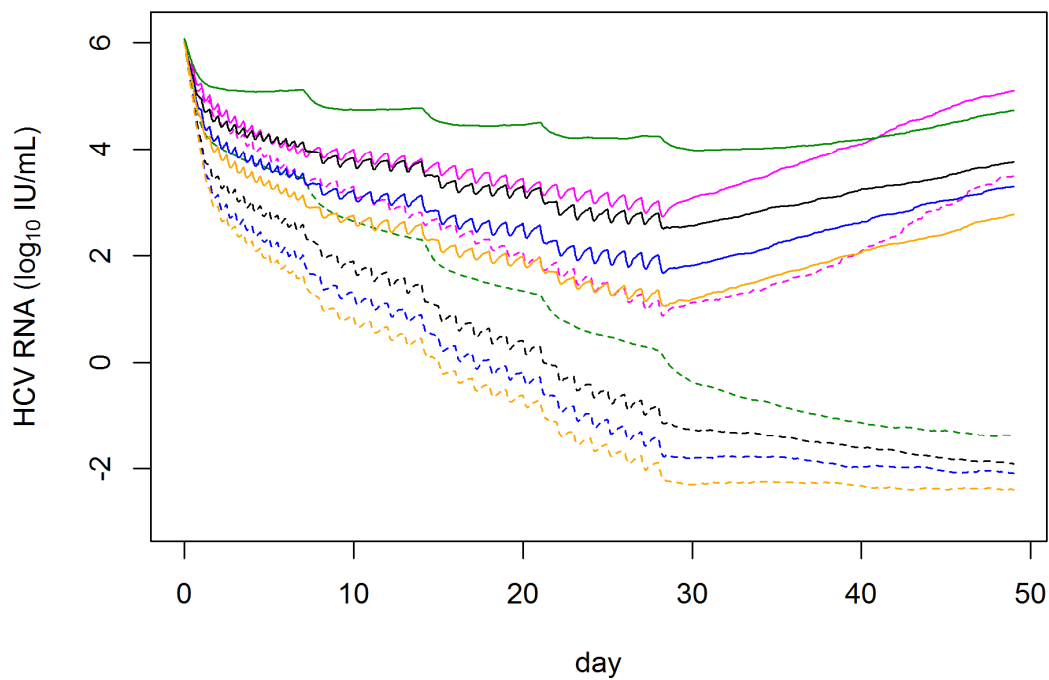

Figure 4. Predicted concentrations of alisporivir and peg-IFN (upper pattern) and predicted virologic response in different treatment arms and in genotype 1/4 and genotype 2/3 patients (lower pattern),

evaluated as the median of 1000 simulations. Simulated profiles are presented in black, blue, orange, magenta, and green for Arm A (alisporivir 200 mg + peg-IFN), B (alisporivir 600 mg + peg-IFN), C (alisporivir 1000 mg + peg-IFN), D (alisporivir 1000 mg) and arm PEG (peg-IFN 180μg), respectively. Virologic responses were presented as solid lines and dashed lines in in genotype-1/4 patients and genotype-2/3 patients, respectively.

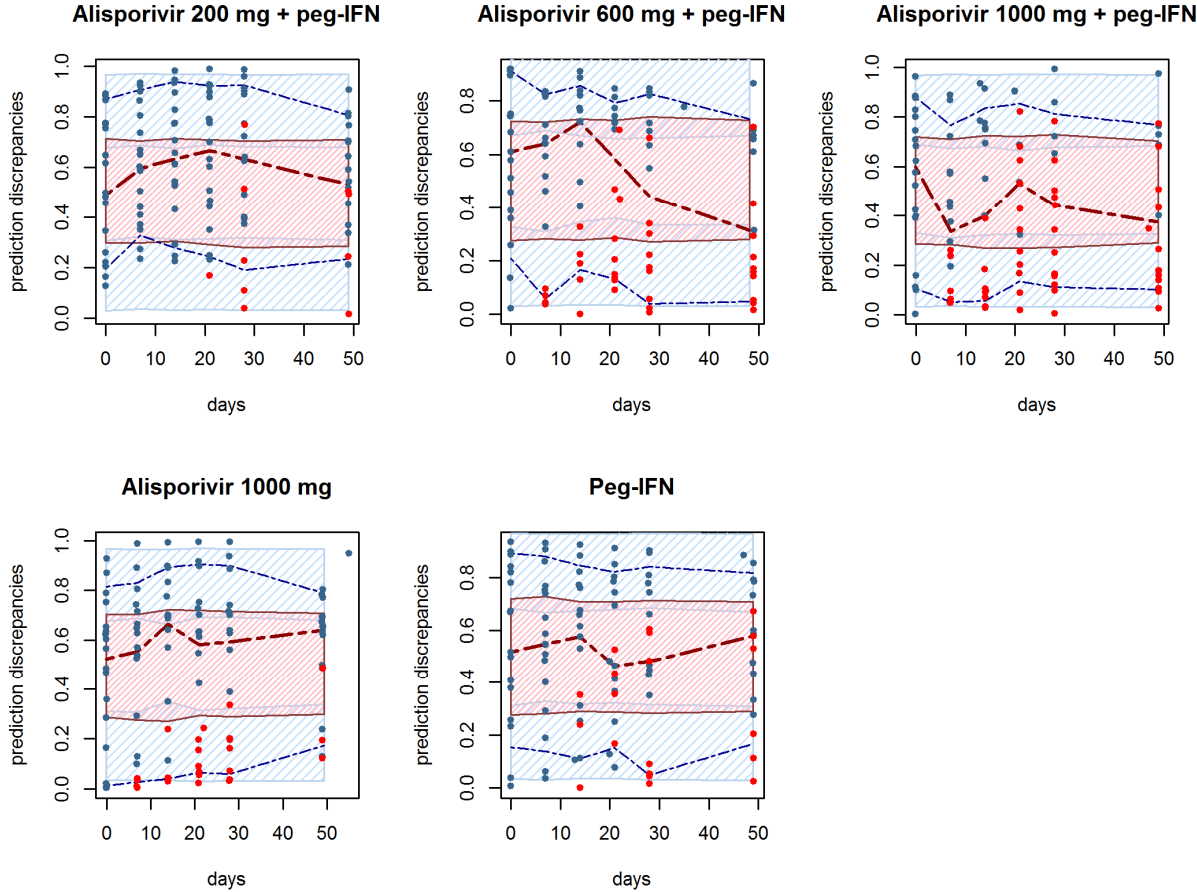

Figure 5. Prediction discrepancies for the PK-VK model for different treatment arms. Dashed lines in blue represent 10$^{th}$ and 90$^{th}$ observed percentiles, dashed line in red represents the 50$^{th}$ observed percentile. Blue and pink shaded areas are the 95% prediction interval of the corresponding percentiles. Observed data are presented in blue; data below the quantification limit are presented in red.

**Table legends**

Table 1. Parameter estimates for PK models of alisporivir, peg-IFN and for the PK-VK model

| Parameters | Fixed effects Estimates (RSE %) | Interindividual variability ω (%) (RSE %) |
|---|---|---|
| **PK parameters of alisporivir** | | |
| $T_{lag}$ (h) | 0.453 (1) | 5.1 (20) |
| $k_a$ (h$^{-1}$) | 2.77 (15) | 46.8 (25) |
| $k_{12}$ (h$^{-1}$) | 0.256 (9) | 26.6 (31) |
| $k_{21}$ (h$^{-1}$) | 0.0419 (12) | 54.6 (16) |
| V (L) | 294 (8) | 36.1 (14) |
| $V_m$ (Peg-IFN = 1) (mg.h$^{-1}$) | 115 (24) | 25.9 (41) |
| $V_m$ (Peg-IFN = 0) (mg.h$^{-1}$) | 55 (24) | |
| $K_m$ (Peg-IFN = 1) (mg.L$^{-1}$) | 2.44 (33) | 39.3 (50) |
| $K_m$ (Peg-IFN = 0) (mg.L$^{-1}$) | 0.64 (11) | |
| $F_0$ | 0.27 | 0 (-) |
| $F_{inf}$ | 0.853 | 97.6 (21) |
| $k_F$ (h$^{-1}$) | 0.0341 | 0 (-) |
| cor(Vm,Km) | - | 0.26 (241) |
| $\sigma_a$ (µg/L) | 6.6 (11) | |
| $\sigma_p$ (%) | 34.1 (4) | |
| **PK parameters of peg-IFN** | | |
| $k_a$ (h$^{-1}$) | 0.0392 (12) | 69.7 (13) |
| V (L) | 12 (5) | 51.4 (11) |
| k (h$^{-1}$) | 0.00603 (7) | 27 (14) |
| $\sigma_a$ (µg/L) | 0.243 (26) | |
| $\sigma_p$ (%) | 30.0 (5) | |
| **VK parameters** | | |
| $VL_0$ (IU.mL$^{-1}$) | $1.14 \times 10^6$ (20) | 166 (9) |
| c (day$^{-1}$) | 6 (-) | 0 (-) |
| δ in GT-1/4 (day$^{-1}$) | 0.217 (8) | 43.1 (9) |
| δ in GT-2/3 (day$^{-1}$) | 0.393 (9) | |
| β (mL.virion$^{-1}$.day$^{-1}$) | $5.6 \times 10^{-7}$ (2) | 171 (13) |
| IC50alv (µg.L$^{-1}$) | 43.4 (38) | 276 (10) |
| IC50peg in GT-1/4 (µg.L$^{-1}$) | 2.26 (43) | 201 (13) |
| IC50peg in GT-2/3 (µg.L$^{-1}$) | 0.146 (54) | |
| $\sigma_a$ (log$_{10}$IU/mL) | 0.366 (4) | |

Table 2. (1) Observed virologic response in VITAL-1 study; (2) Predicted virologic responses obtained by simulation using PK-VK population parameters (including residual errors for VK data). The response-adapted therapy and the number of patients remained in per protocol analysis were taken into account in simulation. Simulation was conducted assuming no ribavirin effect on virologic response and 100% adherence to treatment.

| | (1) | | | | | |
|---|---|---|---|---|---|---|
| | Observed % of patients below LOQ at | | | | Observed SVR (%) | |
| | Week 2 | Week 4 | Week 6 | Week 12 | | |
| VITAL-1b* (N=77) | 22.0 | 37.0 | 50.0 | 99.0 | 91.0 | |
| VITAL-1c* (N = 80) | 20.0 | 42.0 | 45.0 | 98.0 | 91.0 | |
| VITAL-1d* (N=35) | 62.0 | 85.0 | 90.0 | 96.0 | 91.0 | |
| | (2) | | | | | |
| | Predicted % of patients below LOQ** at | | | | Predicted SVR for a 24-week treatment** (%) | |
| | Week 2 | Week 4 | Week 6 | Week 12 | Based on viral load | Based on infected cells |
| VITAL-1b* (N=77) | 19.0 (11.7 – 27.3) | 44.0 (33.8 – 51.9) | 53.6 (42.9 – 63.0) | 92.9 (88.3 - 96.8) | 94.0 (88.3 – 98.7) | 91.7 (85.7 – 96.1) |
| VITAL-1c* (N = 80) | 21.3 (14.3 – 28.8) | 46.8 (36.3 – 58.4) | 57.4 (45.6 – 67.5) | 93.6 (88.1 – 97.5) | 93.6 (88.8 – 98.8) | 92.6 (86.3 – 97.5) |
| VITAL-1d* (N=35) | 61.5 (42.9 – 74.3) | 87.2 (74.3 – 95.8) | 92.3 (84.2 – 100.0) | 96.2 (89.9 - 100.0) | 94.9 (88.6 - 100.0) | 94.9 (87.1 – 100.0) |

*All the three groups of VITAL-1 study received a loading dose of 600 mg alisporivir BID for 1 week, followed by alisporivir 600 mg QD + ribavirin (arm VITAL-1b), alisporivir 800 mg QD + ribavirin (arm VITAL-1c) and alisporivir 600 mg QD + peg-IFN (arm VITAL-1d). Patients with RVR, i.e., HCV RNA below LOQ (<25 IU/mL) at treatment week 4 , continued on the same treatment until week 24, while non-RVR patients switched to alisporivir 600 mg QD plus peg-IFN/ribavirin from week 6 to week 24.
** Median (95% prediction interval)

**Table S1. Patient baseline characteristics per treatment group**

|  | Arm A Alisporivir 200 mg Peg-IFN (N=18) | Arm B Alisporivir 600 mg Peg-IFN (N=18) | Arm C Alisporivir 1000 mg Peg-IFN (N=18) | Arm D Alisporivir 1000 mg (N=18) | Arm Peg Peg-IFN (N=18) |
|---|---|---|---|---|---|
| **Sex (% male)** | 61 | 61 | 72 | 44 | 50 |
| **Age* (years)** | 27.5 (19-51) | 32.0 (19-45) | 40.0 (21-59) | 38.5 (23-68) | 34.5 (21-51) |
| **Body mass index* (kg/m$^2$)** | 24.5 (19-28) | 23.8 (19-28) | 24.4 (19-28) | 24.7 (20-29) | 23.1 (20-29) |
| **Genotypes (GT)** | | | | | |
| 1A | 3 | 4 | 2 | 2 | 1 |
| 1B | 8 | 8 | 10 | 9 | 9 |
| 2A | 0 | 0 | 1 | 1 | 0 |
| 2B | 0 | 0 | 1 | 1 | 1 |
| 3 | 6 | 6 | 4 | 4 | 5 |
| 4 | 1 | 0 | 0 | 1 | 2 |
| **HCV RNA* (log$_{10}$IU/mL)** | | | | | |
| All | 6.03 (5.1-7.1) | 6.40 (4.5–7.4) | 6.27 (4.0–7.6) | 6.12 (3.9–7.2) | 6.04 (4.1–7.2) |
| GT 1 & 4 | 5.80 (5.1-7.1) | 6.46 (5.2–7.4) | 6.45 (4.0–7.0) | 6.23 (4.4–7.2) | 5.93 (4.1–7.2) |
| GT 2 & 3 | 6.5 (5.2-7.0) | 6.20 (4.5–7.2) | 6.02 (5.2–7.6) | 5.82 (3.9–6.7) | 6.71 (5.5–7.1) |
| **Ethnic** | | | | | |
| Caucasian | 18 | 18 | 16 | 17 | 17 |
| Others | 0 | 0 | 2 | 1 | 1 |

*Median (range)

**Table S1. Comparison of structural PK models after last dose of alisporivir**

| Data* | Structural model | -2LL | BIC | Proportional error |
|---|---|---|---|---|
| Day 29 & 50 Arm A, B, C (ALV + Peg-IFN) | Two-compartment model with linear elimination | 4050 | 4093 | 0.32 |
| | Three-compartment model with linear elimination | 4048 | 4106 | 0.32 |
| | Two-compartment model with Michaelis-Menten elimination (Vm,Km) | 4021 | 4071 | 0.31 |
| | **Two-compartment model with Michaelis-Menten elimination and lag time for absorption (model for day 29)** | **3962** | **4020** | **0.25** |
| | Two-compartment model with two eliminations (linear and Michaelis-Menten) and lag time for absorption | 3967 | 4032 | 0.25 |
| | Two-compartment model with Michaelis-Menten elimination, lag time for absorption and nonlinear distribution | 3959 | 4024 | 0.27 |

*333 observations obtained during the last dose (day 29 and 50) from 37 patients in arm A, B, C (ALV 200 mg, 600 mg or 1000 mg + peg-IFN) were used to build the model for the last dose of ALV (for steady state PK)

**Table S3. Comparison of structural PK models of ALVfor alisporivir after first to last dose with changes in PK parameters**

| Data* | Structural model | -2LL | BIC | Proportional error |
|---|---|---|---|---|
| Day 1 - day 50 Arm A, B, C, D | Model of day 29 | 8212 | 8296 | 0.49 |
| | **with time-varying bioavalaibility** | **7863** | **7962** | **0.35** |
| | with time-varying Vm | 8054 | 8159 | 0.41 |
| | with time-varying Km | 8144 | 8249 | 0.45 |

*672 observations from 39 patients with rich design in in arm A, B, C (ALV 200 mg, 600 mg or 1000 mg + peg-IFN) were used to build the structural PK model for ALV

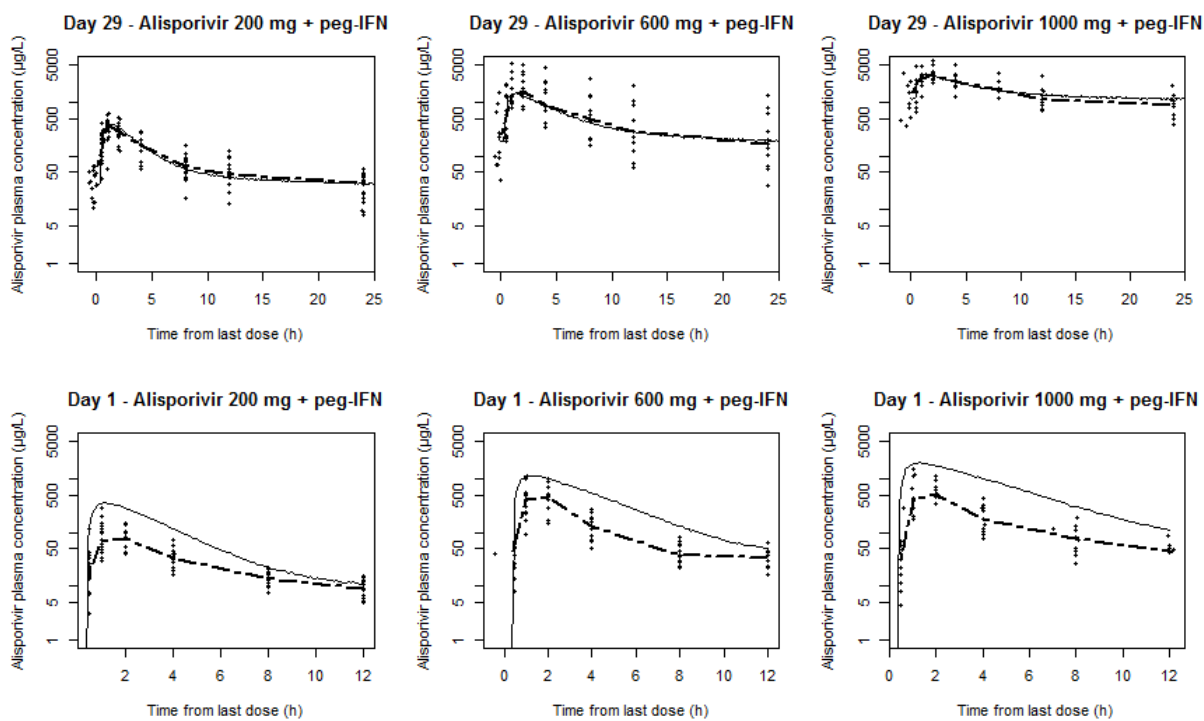

Figure S1. Median of observed data and median predicted by the structural model for the last dose, obtained from 1000 simulations, for the last dose (first row) and for the first dose (last row). This model over-predicted concentration obtained at the first dose of alisporivir

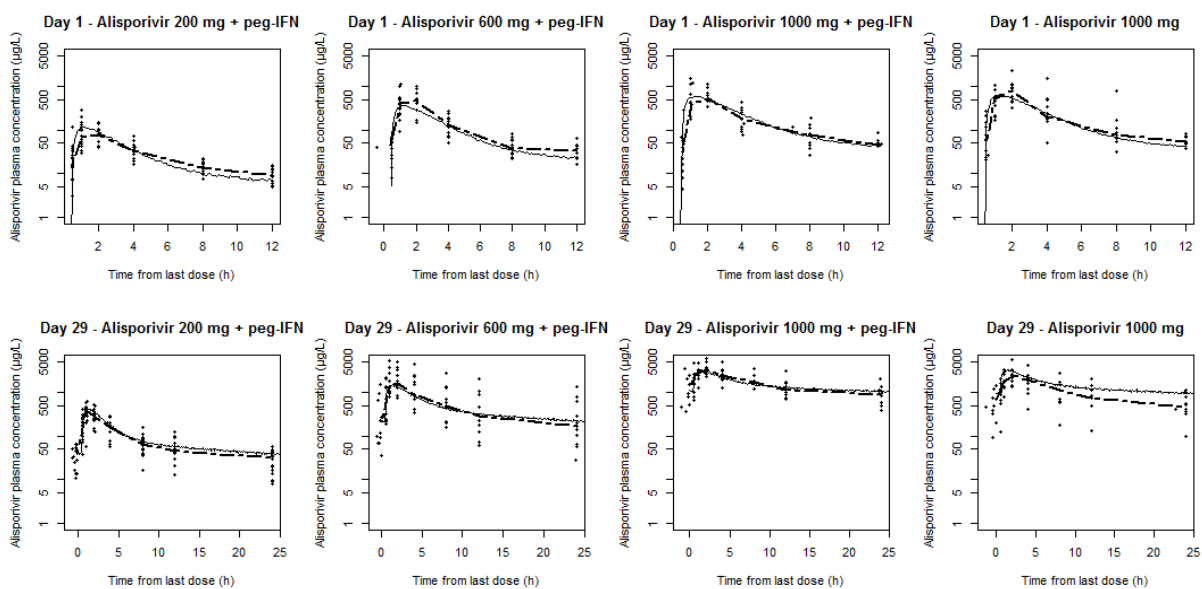

Figure S2. Median of observed data and median predicted by the model for full PK data of three combination groups, obtained from 1000 simulations, for the first dose (first row) and for the last dose (last row) in the four treatment arm. This model over-predicted alisporivir concentration at the last day of treatment in the monotherapy group.

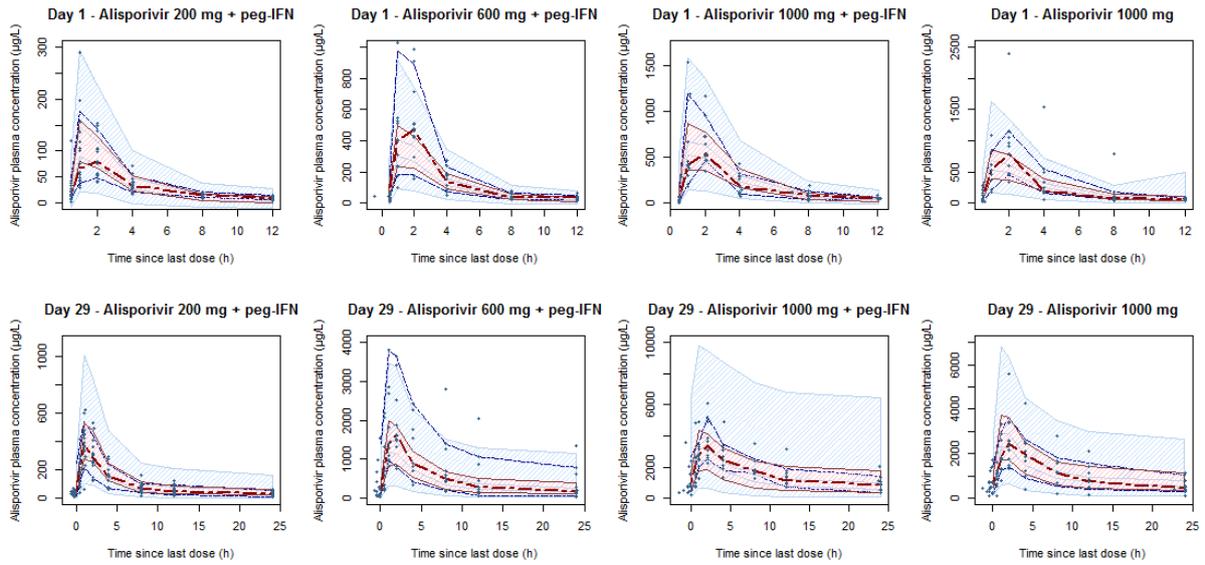

Figure S3. VPC of the full PK model of alisporivir in four treatment arms after taking into account peg-IFN. Dashed lines in blue represent 10th and 90th percentiles of the observations; dashed line in red represents the 50th percentile of the observations. Blue and pink shaded areas are the 95% prediction interval of the corresponding percentiles.

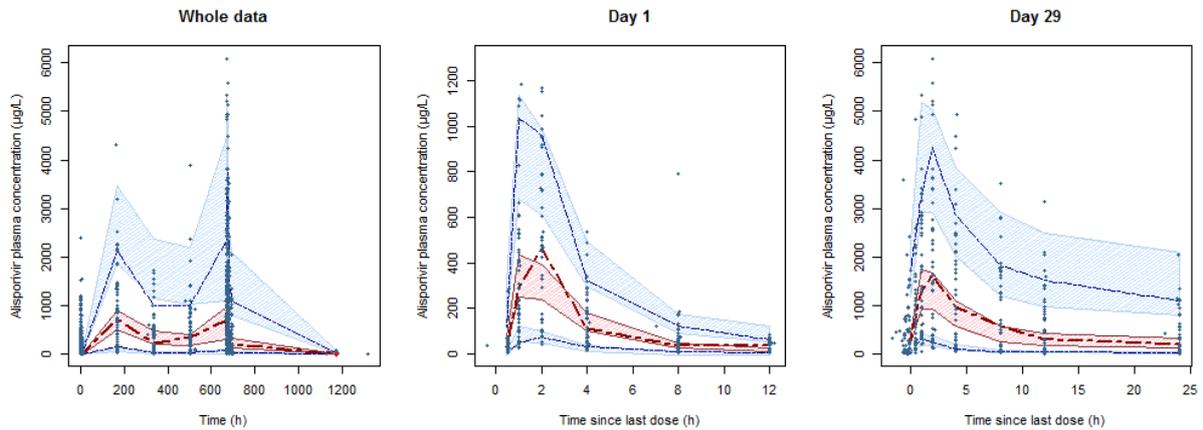

Figure S4. VPC for final alisporivir PK model for all study period (left pattern), day 1 (middle pattern) and the last day of treatment (right pattern). Dashed lines in blue represent 10th and 90th percentiles of the observations; dashed line in red represents the 50th percentile of the observations. Blue and pink shaded areas are the 95% prediction interval of the corresponding percentiles.

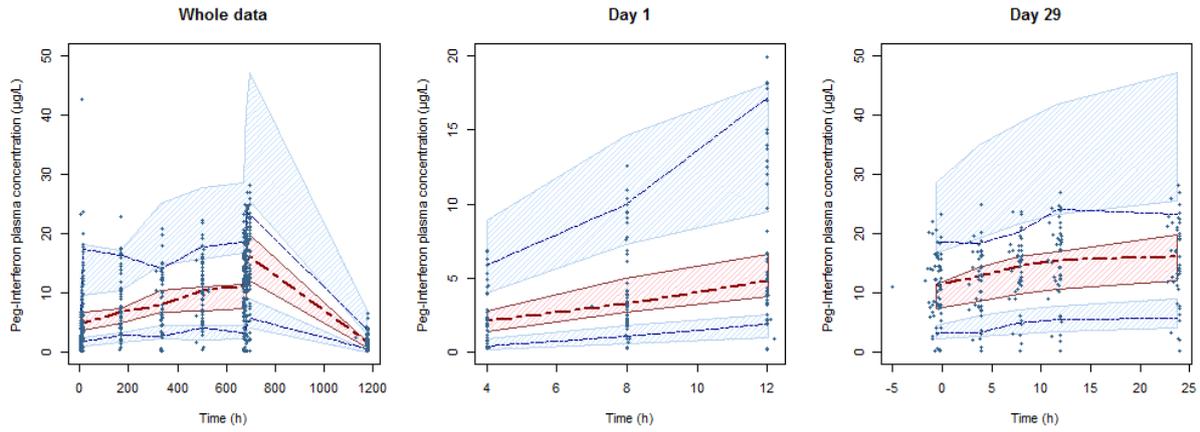

Figure S5. VPC for peg-IFN PK model for all study period (left pattern), day 1 (middle pattern) and the last day of treatment (right pattern). Dashed lines in blue represent 10th and 90th percentiles of the observations; dashed line in red represents the 50th percentile of the observations. Blue and pink shaded areas are the 95% prediction interval of the corresponding percentiles.

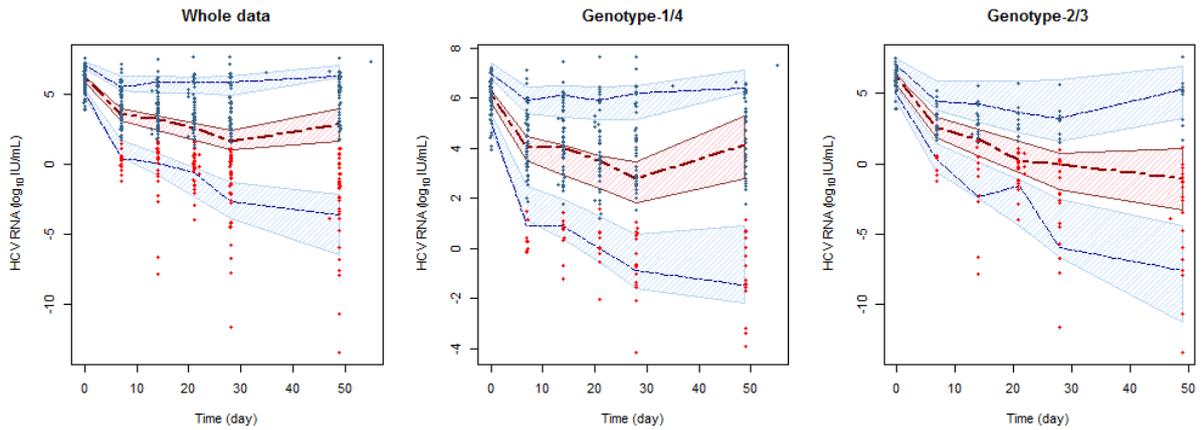

Figure S6. VPC for the PK-VK model for all data (left pattern), for genotype-1/4 patients (middle pattern) and for genotype-2/3 patients (right pattern). Dashed lines in blue represent 10th and 90th percentiles of the observations; dashed line in red represents the 50th percentile of the observations. Blue and pink shaded areas are the 95% prediction interval of the corresponding percentiles. Data below the quantification limit were presented as red closed circle.